# IMPLEMENTASI PENINGKATAN PROFESIONALISME GURU PAUD MELALUI DIKLAT BERJENJANG

Diah Rina Miftakhi[1(a)], Hengki Pramusinto[2(b)]

[1] Universitas Muhammadiyah Bangka Belitung, Indonesia
[2] Universitas Negeri Semarang, Indonesia

a) diahrina1978@gmail.com
b) hpramusinto@mail.unnes.ac.id

| INFORMASI ARTIKEL | ABSTRAK |
|---|---|
| *Kata Kunci:*<br>profesionalisme guru, diklat berjenjang, Pendidikan Anak Usia Dini | *Pelatihan diklat berjenjang merupakan program kegiatan pelatihan yang dilaksanakan secara bertahap. Sasaran kegiatan diklat ini adalah para guru PAUD yang belum mengikuti diklat berjenjang. Penelitian ini bertujuan untuk meningkatkan kemampuan guru dalam mengajar melalui kegiatan pelatihan, yaitu diklat berjenjang. Kegiatan pelatihan ini dilaksanakan secara bertahap. Peserta dinyatakan lulus pada setiap tahapannya apabila telah melaksanakan seluruh kegiatan diklat dan mengumpulkan tugas mandiri kepada pihak penyelenggara diklat. Penelitian ini menggunakan metode penelitian kualitatif deskriptif. Sumber data primer penelitian ini adalah 60 orang guru PAUD di Provinsi Kepulauan Bangka Belitung. Sumber data sekunder penelitian ini adalah hasil tugas mandiri masing-masing peserta, kebijakan tentang pelaksanaan diklat berjenjang, dan artikel jurnal penelitian. Hasil penelitian ini menunjukkan bahwa pelatihan diklat berjenjang yang diikuti dapat berjalan dengan efektif dengan hasil sangat baik serta mampu meningkatkan kemampuan guru dalam mengajar.* |
| | ***ABSTRACT*** |
| *Keywords:*<br>teacher profesionalism, tiered education and training, early childhood education | *Tiered education and training are a program of education and training activities carried out in stages. The target of this training activity is PAUD teachers who have not participated in tiered training activities. This study aims to improve teachers' teaching ability through training activities, namely tiered training. This training activity is carried out in stages. Participants are declared to have passed at each stage if they have carried out all training activities and submitted independent assignments to the education and training organizers. This study uses a qualitative descriptive method. The primary data source for this study was 60 PAUD teachers in Kepulauan Bangka Belitung Province. The secondary data sources of this research are the results of the independent assignments of each participant, policies related to the implementation of tiered training activities, and research journal articles. The results of this study indicate that the following tiered training can run effectively with excellent results and can improve teachers' teaching abilities.* |
| ***Corresponding Author:***<br>E-mail:<br>diahrina1978@gmail.com<br><br>DOI: 10.5281/zenodo.7618997<br>(cc) BY-NC-ND | |

## A. PENDAHULUAN

Guru merupakan sosok pahlawan yang akan memberikan ilmu untuk peserta didik di sekolah. Guru juga merupakan orang yang akan menjadi suri tauladan bagi peserta didik dalam berbagi hal, mulai dari sikap, perilaku, kecerdasan, sampai dalam hal penampilannya. Sehingga seorang guru harus dapat profesional dalam menjalankan tugasnya, yaitu sebagai seorang pendidik. Guru dituntut memiliki profesinalisme yang tinggi untuk melakukan tugas-tugasnya secara profesional. Keterampilan profesional guru dalam proses pengajaran terdiri dari, manajemen kelas yang efektif, keterampilan memotivasi, metode pengajaran, strategi evaluasi, penyusunan rencana pembelajaran dan penggunaan bahan ajar secara efektif (Abusomwan & Osaigbovo, 2020). Guru profesional memiliki peran yang sangat penting yaitu sebagai orang tua kedua bagi peserta didik sehingga guru harus lebih berdedikasi terhadap pekerjaannya secara teratur dengan mulai dari mempersiapkan administrasi pelajaran yang akan dilaksanakan di lembaganya masing-masing (Atma et al, 2021).

Setiap lembaga pendidikan wajib meningkatkan keunggulannya dalam bidang akademik maupun non akademik serta dalam kepemimpinan, staf, pendidik, proses belajar mengajar, pengembangan pendidik dan tenaga kependidikan, kurikulum, iklim sekolah, dan keterlibatan orang tua dan masyarakat (Direktorat Jendral Guru dan Tenaga Kependidikan, 2018). Tingkat keberhasilan lembaga pendidikan dalam penyelenggaraan pendidikannya sangat ditentukan oleh keberhasilan guru pada proses pembelajaran (Safitri, Hadiyanto & Ramli: 2018). Selain itu tingkat kualifikasi akademik guru juga akan mempengaruhi tingkat kompetensi dan kemampuannya dalam mengajar. Guru Pendidikan Anak Usia Dini (PAUD) harus memenuhi kualifikasi dan kompetensi yang dipersyaratkan sesuai dengan Permendikbud No. 137 Tahun 2014 tentang Standar Nasional PAUD Pasal 24 yang menyatakan bahwa pendidik anak usia dini terdiri dari tiga kategori, yakni guru PAUD, guru pendamping, dan guru pendamping muda.

Hasil penelitian Yusutria (2017), menjelaskan bahwa guru yang profesional itu tercermin keberhasilannya dalam melaksanakan tugas mengajarnya ditandai oleh penguasannya terhadap materi baik teori maupun prakteknya. Guru profesional dapat dipengaruhi oleh: a) kualifikasi pendidikan, b) ikut serta di berbagai pengembangan diri (pelatihan, penyetaraan, workshop serta berbagai penataran yang telah diikuti sesuai bidangnya, c) menjalin interaksi yang padu dengan berbagai elemen masyarakat, d) pengembangan kinerja yang mencerminkan mutu dan pelayanan prima, e) menguasai dan mengintegrasikan teknologi dan komunikasi. Sehingga jelas sekali bahwa seorang guru harus profesional dalam mengajar dan harus memiliki kemampuan kompetensi yang baik.

Guru profesional harus mampu mengenal peserta didiknya. Hal ini seperti yang disampaikan oleh Fahruddin & Nilawati (2018), guru profesional harus berkompeten menyiapkan materi pembelajaran, kemampuan menyelenggarakan pembelajaran mulai dari perancangan hingga evaluasi dan pemanfaatan hasil evaluasi untuk mengembangkan kemampuan profesionalnya secara berkelanjutan. Demikian juga dengan kompetensi guru untuk Pendidikan Anak Usia Dini (PAUD). Profesionalisme guru PAUD dapat dilihat dari kualifikasi pendidikan dan pengalaman mengajar mereka. Ketercapaian kualifikasi pendidikan akan membuat seorang pendidik memenuhi ketentuan dan standar PAUD, sehingga memiliki kompetensi yang sesuai dan dibutuhkan dalam mengelola pembelajaranan anak usia dini (Priansa & Somad, 2014). Faktor penentu kualitas pendidikan adalah guru profesional. Rendahnya kualitas pendidikan menandakan minimnya keberadaan guru professional (Kristiawan & Rahmat, 2018). Sehingga penting sekali bahwa guru PAUD dituntut untuk lebih profesional dalam mengajar, karena guru PAUD orang yang pertama kali mengajarkan dasar-dasar pendidikan kepada anak.

Layanan PAUD merupakan langkah awal bangsa untuk menciptakan generasi yang mampu bersaing dengan negara-negara lain. (Rozalena dan Kristiawan, 2017). Disampaikan oleh Direktorat PAUD (2015) bahwa berbagai penelitian yang terkait PAUD menunjukkan adanya penyiapan SDM berkualitas harus diawali sejak dini, bahkan sejak masih masa konsepsi dalam kandungan. Mengetahui betapa pentingnya PAUD, maka Guru PAUD sangat berperan dalam memberikan pendidikan. Peran guru yang awalnya hanya penyuap pengetahuan, berkembang menjadi mentor, fasilitator, motivator, inspiratory, juga pengembang imajinasi dan kreativitas. Guru menanamkan nilai karakter dan membangun kerjasama dan empati sosial. Aspek-aspek yang demikian menjadi teramat penting dijalankan oleh guru karena peran seorang takkan tergantikan bahkan dengan mesin sekalipun. Selain itu saat ini mulai berkembang masalah-masalah dalam dunia pendidikan, sehingga peran seorang guru profesional harus ditingkatkan lagi.

Persoalan pendidikan yang banyak terjadi sampai saat ini berkaitan dengan mutu proses pembelajaran yang rendah, pemerintah yang belum berkomitmen sepenuhnya untuk membangun pendidikan dan meningkatkan mutu SDM, kurikulum yang terus berganti dan tidak terealisasi dengan baik, lemahnya kompetensi guru. Pemerintah haruslah berupaya membenahi pendidikan, melakukan intervensi serta berbagai bentuk dukungan dalam mempercepat proses pembangunan. (Ramdhani & Santosa, 2012).

Permasalahan mendasar dalam pendidikan adalah permasalahan dalam kualitas SDM, sehingga sangat dibutuhkan upaya peningkatan kemampuan SDM khususnya tenaga pendidik yang belum optimal dalam mengembangkan potensi-potensi yang dimilikinya. (Zakiya & Nurhafizah, 2019). Beberapa hasil penelitian sebelumnya juga menunjukkan bahwa profesionalisme guru Indonesia masih harus ditingkatkan (Siswandari, 2013; Lestari & Purwanti, 2018; Rohmat, 2019). Hal ini karena guru sebagai pihak yang terlibat langsung dalam pembelajaran di kelas, memiliki peran yang sangat vital dalam peningkatan kualitas peserta didiknya. Keberhasilan proses pendidikan dapat dikatakan sangat tergantung pada peran guru di sekolah. Selain itu guru menentukan keberhasilan pendidikan (Dwirahayu, et al, 2020) dan merupakan penentu untuk faktor eksternal yang akan mempengaruhi mutu pendidikan. (Dahar 2011; Syarifuddin, 2011).

Selain itu permasalahan yang lain juga terjadi dalam pendidikan usia dini. Kondisi nyata menunjukkan bahwa Guru PAUD, terutama di jalur nonformal memiliki variasi yang beragam secara kualifikasi maupun kompetensi.

Di lapangan masih banyak ditemui guru dengan pendidikan SMA/SMK bahkan masih dijumpai guru yang berpendidikan SD/SMP. Meskipun sebagian telah mempunyai kualifikasi diploma/sarjana, namun seringkali tidak sesuai dengan bidang PAUD. (https://disdik.bogorkab.go.id/post/upaya-meningkatkan-kompetensi-guru-paud-melalui-diklat-berjenjang, 2019). Pendapat lain juga disampaikan oleh Nofriyanti dan Nurhafizah (2019), bahwa salah satu kendala kualitas guru yang masih rendah disebabkan oleh masih adanya guru-guru PAUD yang mengajar di lembaga PAUD tetapi bukan dari program studi atau jurusan PAUD. Sehingga kualifikasi akademik seorang Pendidik Anak Usia Dini harus lebih diperhatikan serta ditingkatkan lagi.

Kondisi yang telah dipaparkan memperkuat alasan perlunya kegiatan Pendidikan dan Pelatihan (diklat) bagi para guru PAUD secara berkesinambungan sesuai dengan jenjang keguruannya. Diklat tersebut meliputi diklat dasar, diklat lanjut, dan diklat mahir. Diklat dasar sangat penting dalam menyiapkan kompetensi minimal guru sebagai guru pendamping muda. Diklat lanjut berperan dalam menyiapkan kompetensi minimal sebagai guru pendamping. Diklat mahir sangat penting diikuti untuk mencapai kompetensi minimum sebagai guru PAUD.

Diklat berjenjang ini memiliki tujuan yaitu untuk mengasah kemampuan guru PAUD. Diklat ini dilakukan supaya guru PAUD tidak sampai melakukan malpraktek dalam mendidik Anak Usia Dini. Sebab apabila terjadi malpraktek dalam pendidikan dampaknya tidak akan langsung terlihat saat ini, tetapi dampak malpraktek tersebut baru akan terlihat beberapa tahun kemudian pada saat anak sudah menempuh pendidikan lebih tinggi. (Bangkapost.com, 2019). Kegiatan pelatihan diklat berjenjang yang dilaksanakan untuk guru-guru PAUD ini diharapkan akan dapat membantu meningkatkan kemampuan guru dalam mengajar mulai dari persiapan administrasi sebelum proses pembelajaran sampai pada proses evaluasinya sehingga guru dapat melaksanakan proses pembelajaran menjadi lebih baik lagi dan lebih profesional. Seperti yang jelaskan oleh Gutierez (2020) bahwa kemampuan menyusun adminsitrasi pembelajaran merupakan ciri guru professional selanjutnya menerapkannya untuk meningkatkan kompetensi siswa (Stemberger, 2020), dan mengukur hasil kemajuan pembelajaran siswa (De-Simone, 2020).

Pelaksanaan diklat berjenjang masih sangat dibutuhkan oleh guru-guru PAUD untuk meningkatkan kompetensinya dalam mengajar. Penyelenggara diklat berjenjang dapat dilaksanakan mulai provinsi sampai tingkat kabupaten/kota yang sudah memiliki persyaratan untuk menyelenggarakan kegiatan diklat tersebut. (Kemendiknas, 2011). Guru-guru yang sudah mengikuti diklat berjenjang ini apabila sudah selesai dalam mengikuti kegiatan diklat maka akan diberikan tugas mandiri. Setelah tugas mandiri selesai dikerjakan dan sudah lengkap semuanya, guru PAUD baru akan dinyatakan lulus dan selanjutnya akan mendapatkan sertifikat. Diharapkan diklat berjenjang ini akan terus dilaksanakan oleh semua pihak terutama dari Dinas Pendidikan masing-masing Kabupaten/Kota yang ada di Provinsi Kepulauan Bangka Belitung atau diklat berjenjang yang akan dilaksanakan secara mandiri.

Berdasarkan beberapa permasalahan yang muncul tentang kurangnya kompetensi guru PAUD serta masih perlunya peningkatan kemampuan profesionalisme guru dalam menyiapkan kegiatan dalam proses pembelajaran, maka diklat berjenjang untuk guru-guru PAUD ini masih sangat penting untuk dilaksanakan dan masih harus dilakukan evaluasi untuk dapat meningkatkan kemampuan dan kapabilitas guru PAUD, khususnya yang ada di wilayah Provinsi Kepulauan Bangka Belitung.

## B. Metode Penelitian

Penelitian ini menggunakan jenis penelitian deskriptif kualitatif. Jenis penelitian deskriptif kualtitatif digunakan untuk melakukan penelitian pada kondisi obyek yang alamiah (Sugiyono, 2019: 11). Sesuai dengan jenis penelitian yang digunakan, pendekatan penelitian yang dipilih peneliti adalah pendekatan deskriptif. Pendekatan deskriptif mendasarkan pertimbangan bahwa permasalahan yang hendak diteliti sedang berlangsung pada masa sekarang dengan tujuan menganalisa fenomena-fenomena yang terjadi di lapangan.

Teknik pengumpulan data menggunakan teknik observasi, teknik wawancara, dan teknik dokumentasi (Hendrik, 2018). Berdasarkan teknik pengumpulan data tersebut, instrumen penelitian ini adalah tes (pre-test dan post-test), angket, pedoman wawancara. Data yang terkumpul kemudian dianalisis secara deskriptif dengan menggunakan teknik reduksi data, display data, dan penarikan simpulan (Miles & Hubberman,1992).

Reduksi data yaitu suatu proses pemilihan, pemusatan perhatian pada penyederhanaan, pengabstrakan dan transformasi data kasar yang muncul dari catatan-catatan tertulis di lapangan. Reduksi data yang berupa transkrip hasil wawancara terhadap subjek penelitian. Setelah data direduksi, data disusun sedemikian rupa sehingga memberikan kemungkinan adanya penarikan kesimpulan dan pengambilan tindakan. (Hendrik, 2018). Dalam penelitian ini akan diperoleh kesimpulan yang tentatif, kabur, kaku dan meragukan, sehingga kesimpulan tersebut perlu diverifikasi. Verifikasi dilakukan dengan melihat kembali reduksi data maupun display data sehingga simpulan yang diambil tidak menyimpang.

Penelitian ini dilakukan di Provinsi Kepulauan Bangka Belitung. Sumber data penelitian ini yaitu sumber data primer dan sekunder. Sumber data primer meliputi guru-guru PAUD di Provinsi Kep. Bangka Belitung dengan total peserta sebanyak 60 orang, dan akan dibagi kedalam tiga kelas yaitu 20 peserta dikelas diklat tingkat dasar, 20 peserta di kelas diklat tingkat lanjut, serta 20 peserta di kelas diklat tingkat mahir. Sedangkan sumber data sekunder adalah hasil tugas mandiri yang dilaksanakan oleh masing-masing peserta pelatihan, kebijakan pemerintah, dan berbagai literatur yang berasal dari artikel jurnal hasil penelitian.

## C. Hasil Penelitian

Pelayanan pendidikan bermutu sangat penting dan sangat dibutuhkan oleh semua masyarakat. Guru mempunyai tanggungjawab besar terhadap keberhasilan proses pembelajaran. Proses yang baik akan menghasilkan pengetahuan dan ketrampilan yang akan digunakan oleh peserta didik dalam meraih masa depannya dan dapat diterapkan dalam hidup bermasyarakat. (Nurhafizah, 2018). Tugas dan tanggungjawab guru dalam pembelajaran merupakan bentuk kemampuan profesionalnya.

Sehingga upaya peningkatan profesionalitas guru perlu dilakukan melalui peningkatan kemampuan dan kompetensi guru. Peningkatan yang dimaksud berkaitan dengan usaha membina para guru agar pengembangan kualitasnya tercapai. Usaha yang dilakukan dengan cara melakukan diklat, seminar, workshop, dan pelatihan lainnya yang terkait bidang tugas pendidik. Pengembangan variasi bentuk kecerdasan dalam pembelajaran membutuhkan kemampuan pengetahuan yang memadai, yang dimiliki oleh guru untuk mencapai tujuan pembelajaran dan kemampuan serta potensi peserta didik dapat berkembang maksimal. (Nurhafizah, 2017).

Salah satu kegiatan yang dilaksanakan untuk meningkatkan kompetensi guru-guru PAUD di Provinsi Kepuluan Bangka Belitung yaitu diklat berjenjang. Kegiatan ini dilaksanakan dengan harapan semua guru-guru PAUD dapat mengikuti diklat berjenjang dengan baik, serta ilmu dan pengalaman yang sudah diperoleh selama mengikuti diklat berjenjang ini dapat langsung diimplementasikan di lembaganya masing-masing supaya dapat meningkatkan mutu sekolah.

Diklat berjenjang yang dilaksanakan meliputi diklat tingkat dasar, diklat tingkat lanjutan, dan diklat tingkat mahir. Pelaksanaan diklat berjenjang ini dilaksanakan dengan beberapa rangkaian kegiatan yang harus diikuti oleh semua peserta diklat. Kegiatan tersebut mulai dari pemberian materi oleh narasumber yang profesional dan kegiatan yang berupa penugasan secara mandiri oleh masing-masing peserta diklat.

Peserta dapat dinyatakan lulus diklat berjenjang ini salah satunya yaitu harus mengumpulkan laporan hasil tugas mandiri yang sudah disetujui hasil laporannya oleh pendamping tugas mandiri di lapangan.

**Tabel 1. Rerata Penilaian Tugas Mandiri**

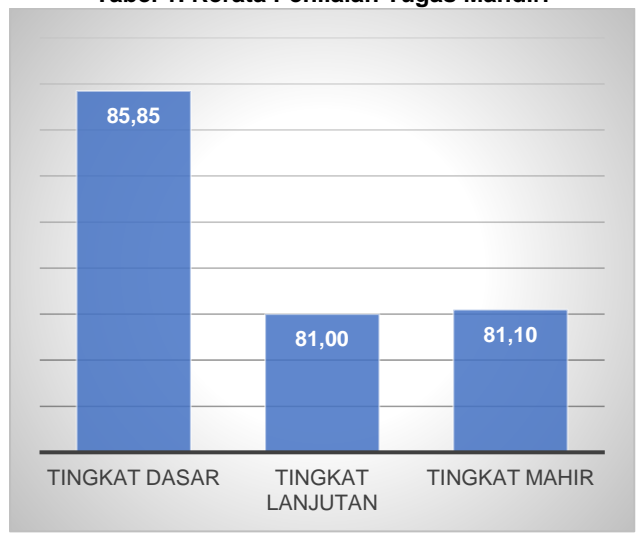

Berdasarkan Tabel 1 dapat dijelaskan, bahwa dengan dilaksanakan diklat berjenjang ini, maka tingkat pemahaman dan kompetensi guru-guru PAUD dapat lebih meningkat dengan baik. Hal ini dapat dilihat dari hasil evaluasi selama proses kegiatan diklat yang sudah dilaksanakan serta hasil evaluasi tugas mandiri yang sudah dilaksanakan oleh seluruh peserta. Tugas mandiri dapat diselesaikan dengan predikat sangat baik dan baik, serta guru-guru PAUD dapat langsung mempraktikkan pengetahuan dan ketrampilan yang mereka peroleh selama kegiatan diklat dilembaganya masing-masing dalam proses kegiatan pembelajaran. Praktik yang dilakukan tersebut mulai dari tahap persiapan mengajar, menyiapkan media pembelajaran dan bahan untuk mengajar, sampai pada hasil evaluasi mengajar yang berupa penilaian terhadap hasil belajar peserta didik. Sehingga dapat disimpulkan bahwa, diklat berjenjang ini sangat bermanfaat sekali untuk guru-guru PAUD untuk dapat menambah pengalaman mereka dalam praktik mengajar.

### D. PEMBAHASAN

Pendidikan Anak Usia Dini merupakan program prioritas pertama dan utama dalam fokus pembangunan pendidikan Indonesia tahun 2010-2014 (Kemendiknas, 2011). Peran pendidik PAUD nyata terlihat dalam keberhasilannya merawat, mengasuh, mendidik, dan melindungi anak sebagai upaya memaksimalkan terkoneksinya seluruh sel otak yang saat lahir sudah terbentuk. Pendidik PAUD adalah tenaga profesional yang memiliki tugas komprehensif, dimulai dari perencanaan, pelaksanaan, sampai pada menilai hasil pembelajaran, serta mengasuh, membimbing, dan melindungi anak didik. Oleh karena itu menjadi keharusan bagi pendidik PAUD untuk memenuhi kualifikasi dan kompetensinya sebelum melaksanakan tugas dan kewajibannya sebagai guru PAUD.

Guru merupakan seorang pendidik profesional dengan tugas yang komplit dan komprehensif. Pengajaran, pembimbingan, pengarahan, pelatihan, dan penilaian peserta didik menjadi bagian yang tidak terpisahkan dari tugas seorang guru. Profesionalisme guru merupakan kondisi, arah, nilai, tujuan dan kualitas suatu keahlian dan wewenang dalam bidang pendidikan dan pengajaran yang berkaitan dengan pekerjaan seseorang yang menjadi tugas pokoknya. Tenaga pendidik PAUD harus selalu meningkatkan kemampuan dan kompetensinya dalam mengajar, hal ini karena guru tidak hanya memiliki tugas untuk mentransfer ilmu pengetahuan, keterampilan dan teknologi saja, melainkan juga harus mengemban tugas yang dibebankan masyarakat kepadanya. Tugas tersebut meliputi mentransfer kebudayaan dalam arti luas, keterampilan dalam menjalani hidup (*life skills*), dan nilai serta *beliefs* untuk anak-anak usia dini.

Kegiatan pembelajaran yang baik untuk memberikan layanan anak usia dini dapat diwujudkan apabila guru dan peserta didik tidak membatasi diri dalam berkomunikasi dalam melaksanakan kegiatan pembelajaran. Hubungan yang harmonis antara guru dan peserta didik akan membuat peserta didik tidak takut dan ragu dalam mengungkapkan masalah belajarnya (Suprihatiningrum, 2012). Hubungan yang demikian hanya dapat tercipta apabila seorang guru memiliki kemampuan berkomunikasi yang baik. Kompetensi guru merupakan kemampuan guru untuk mentransfer pengetahuan dan keterampilannya dalam melaksanakan kewajiban pembelajaran secara profesional dan bertanggung jawab. Saragih (2008) menyatakan bahwa kompetensi minimal seorang guru adalah menguasai keterampilan mengajar dalam hal membuka dan menutup pelajaran, bertanya, memberi penguatan, dan mengadakan variasi mengajar untuk peserta didiknya.

Keberhasilan penyelenggaraan PAUD sejalan dengan peran para pendidik PAUD. Hal ini seperti yang terdapat dalam Undang-Undang Nomor 20 Tahun 2003 tentang Sistem Pendidikan Nasional, pasal 28 disebutkan bahwa keberhasilan dalam menyelenggarakan PAUD sangat tergantung dari peran seorang guru. PAUD dapat diselenggarakan melalui jalur pendidikan formal, nonformal, dan informal. Pendidikan, pengajaran, pembimbingan, pengarahan, pelatihan, dan penilaian peserta didik merupakan tugas utama seorang pendidik. Untuk dapat menjalankan tugas utama tersebut seorang pendidik harus memiliki kompetensi pedagogik, kompetensi kepribadian, kompetensi sosial, dan kompetensi profesional.

Perlunya Pendidikan dan Pelatihan (diklat) bagi para guru PAUD secara berkesinambungan sesuai dengan jenjang keguruannya agar tercapai guru profesional. Diklat tersebut meliputi diklat dasar, lanjut, dan mahir. Diklat dasar sangat penting dalam menyiapkan kompetensi minimal guru sebagai guru pendamping muda. Diklat lanjut berperan dalam menyiapkan guru pendamping untuk menguasai kompetensi minimalnya. Diklat mahir sangat penting diikuti untuk mencapai kompetensi minimum sebagai guru PAUD

Kegiatan diklat dilakukan mulai dari tingkat pusat, provinsi sampai kabupaten/kota. Kegiatan ini hendaknya diselenggarakan oleh lembaga yang memenuhi syarat atau telah ditetapkan sebagai penyelenggara. Keberhasilannya ditentukan oleh faktor pelatih, kualitas pendamping, pengevaluasi, dan strategi pelaksanaan diklat serta komponen yang lainnya.

Diklat berjenjang memiliki tujuan berbeda pada setiap jenjangnya. Diklat berjenjang tingkat dasar memiliki tujuan untuk mempersiapkan pendidik agar mencapai kompetensi: 1) pemahaman terhadap konsep dasar materi PAUD, 2) pemahaman dan penerapan etika dan karakter pendidik PAUD, 3) pemahaman dan penerapan perkembangan dan cara belajar anak usia dini, 4) pemahaman dan penerapan pengenalan anak yang mempunyai kebutuhan khusus, 5) pemahaman dan penerapan materi kesehatan dan gizi anak dini usia, 6) pemahaman dan penerapan materi perencanaan pembelajaran, 7) pemahaman dan penerapan materi penilaian perkembangan anak dini usia, dan 8) pemahaman dan penerapan materi komunikasi dalam pengasuhan.

Diklat berjenjang tingkat lanjutan memiliki tujuan untuk mempersiapkan pendidik yang memiliki kompetensi: 1) pemahaman dan penerapan materi kurikulum PAUD, 2) pemahaman dan penerapan materi strategi pembelajaran 6 aspek perkembangan anak usia dini, 3) pemahaman dan penerapan materi anak berkebutuhan khusus dan cara belajarnya, 4) pemahaman dan penerapan materi deteksi tumbuh kembang anak usia dini, 5) pemahaman dan penerapan materi perencanaan pembelajaran komprehensif, 6) pemahaman dan penerapan materi penilaian perkembangan dan belajar anak usia dini, dan 7) materi strategi pelibatan orangtua, keluarga dan masyarakat dalam PAUD.

Diklat berjenjang tingkat mahir memiliki tujuan untuk mempersiapkan pendidik yang memiliki kompetensi dalam hal: 1) pemahaman dan penerapan materi pengelolaan kurikulum PAUD, 2) pemahaman dan penerapan materi pendidikan inklusif dalam PAUD, 3) pemahaman dan penerapan materi teori belajar dan prinsip-prinsip pembelajaran anak dini usia, 4) pemahaman dan penerapan materi intervensi dini tumbuh kembang anak, 5) pemahaman dan penerapan materi rencana pembelajaran secara inovatif, 6) pemahaman dan penerapan materi penelitian tindakan kelas, 7) pemahaman dan penerapan materi kepribadian multikulturalisme, 8) pemahaman dan penerapan materi pengembangan PAUD holistik integratif, 9) pemahaman dan penerapan materi pemanfaatan teknologi informasi dalam pembelajaran anak usia dini.

Seluruh peserta diklat sebelum mengikuti diklat berjenjang wajib untuk mengikuti kegiatan pretest terlebih dahulu. Kegiatan pretest ini dilakukan untuk mengetahui sejauh mana tingkat pemahaman masing-masing peserta terhadap materi-materi yang akan di sampaikan oleh narasumber. Dari hasil pretest tersebut akan digunakan oleh panitia sebagai bahan diskusi yang akan disampaikan kepada narasumber, sehingga para narasumber bisa memberikan materi dengan menggunkan metode dan bahan ajar yang tepat. Selain itu setiap narasumber dalam menyampaikan materinya juga sudah ditentukan jadwal dan waktu pelaksanaanya supaya materi yang disampaikan dapat tersampaikan semuanya kepada seluruh peserta diklat. Hal ini dilakukan supaya materi yang disampaikan oleh para narasumber dapat dipahami oleh peserta dan setelah kegiatan pelatihan selesai, maka guru-guru dapat mengimplementasikan ilmu-ilmu tersebut dilembaganya masing-masing.

Setelah kegiatan pretest dan penyampaian materi selesai dilaksanakan, maka kegiatan selanjutnya yaitu seluruh peserta akan diberikan penjelasan tentang tugas mandiri yang harus dilaksanakan oleh masing-masing peserta. Tugas mandiri ini sifatnya wajib dan dikerjakan secara individu. Tugas mandiri tersebut merupakan salah satu bentuk pengimplementasian materi-materi yang sudah diperoleh oleh peserta selama mengikuti kegiatan diklat. Dalam tugas mendiri tersebut peserta disuruh untuk melakukan praktik mengajar mulai dari menyusun rencana kegiatan pembelajaran, menyiapkan bahan dan media yang akan digunakan untuk mengajar, melakukan proses pembelajaran, sampai pada tahap melakukan evaluasi dari hasil pembelajaran tersebut.

Bentuk evaluasi pembelajaran dilaksanakan selama kegiatan tugas mandiri. Tugas mandiri dilaksanakan langsung di lembaganya masing-masing. Seluruh tugas-tugas mandiri harus diselesaikan dengan baik dan selama penyelesaian tugas mandiri tersebut, masing-masing peserta akan dibagi menjadi beberapa kelompok. Setiap kelompok akan didampingi oleh masing-masing pendamping sebagai evaluator untuk melakukan evaluasi terhadap hasil tugas mandiri. Pendampingan akan dilakukan selama beberapa kali pertemuan, dan setiap pertemuan akan dilihat hasil perkembangan penyusunan laporan tugas mandiri masing-masing peserta.

Apabila semua laporan tugas mandiri tersebut dinyatakan lengkap dan benar serta sudah sesuai dengan pedoman penulisan laporan tugas mandiri untuk masing-masing diklat, maka tugas mandiri sudah dapat dikumpulkan kepada masing-masing evaluator untuk diberikan penilaian sebagai tugas akhir dalam mengikuti diklat berjenjang tersebut. Kemudian apabila masih ada peserta yang hasil laporan tugas mandirinya masing belum lengkap, maka evaluator akan mengembalikan tugas mandiri tersebut kepada peserta untuk dilakukan perbaikan atau dilakukan revisi, setelah dilakukan revisi, tugas mandiri akan diperiksa lagi oleh evaluator dan diberikan penilaian. Untuk peserta yang sudah lengkap semua tugas-tugasnya serta laporan mandirinya juga sudah lengkap, maka peserta akan dinyatakan lulus dan akan diberikan sertifikat.

Sertifikat yang diberikan kepada peserta sesuai dengan jenis diklat yang diikuti oleh guru PAUD. Bagi yang sudah lulus dan mendapatkan sertifikat diklat tingat dasar, guru PAUD tersebut dapat mengikuti diklat selanjutnya yaitu diklat berjenjang tingkat lanjutan. Untuk guru PAUD yang sudah lulus dan mendapatkan sertifikat diklat berjenjang tingkat lanjutan, maka dapat mengikuti diklat selanjutnya yaitu diklat berjenjang tingkat mahir.

Berdasarkan hasil diklat berjenjang yang sudah telah dilaksanakan maka dapat diketahui bahwa diklat berjenjang ini sangat efektif untuk dilaksanakan. Hasil yang diperoleh menunjukkan bahwa semua peserta dapat mengikuti kegiatan tersebut dengan sangat baik. Seluruh peserta dapat mengikuti semua kegiatan dengan sangat baik mulai dari kegiatan pemberian materi dari masing-masing narasumber dan seluruh peserta dapat mengumpulkan tugas mandiri dengan hasil nilai yang masuk dalam kategori baik dan baik sekali. Dari hasil pelaksanaan diklat berjenjang tersebut para guru-guru PAUD juga dapat langsung mengimplementasikan ilmu-ilmu dan ketrampilan yang mereka peroleh selama kegiatan diklat tersebut dilaksanakan.

Selain itu guru-guru PAUD juga langsung dapat melakukan perbaikan terhadap proses kegiatan pembelajaran yang selama ini sudah mereka laksanakan. Dalam tugas mandiri yang dilaksanakan para peserta diklat dapat langsung melakukan pengamatan, membuat media belajar, menyusun rencana pembelajaran dan dapat langsung melakukan penilaian terhadap hasil mengajar. Sehingga dari hasil penilaian tugas mandiri tersebut dapat diketahui bahwa, guru mengalami peningkatan kompetensinya dan guru juga dapat menjadi lebih professional dalam menjalankan tugasnya.

## E. PENUTUP

Sebagai tenaga profesional, pendidik PAUD harus mempunyai seperangkat kompetensi yang komprehensif dalam melaksanakan tugas dan tanggungjawabnya. Mulai dari tahap perencanaan hingga mampu menggunakan hasil evaluasi pembelajaran untuk meningkatkan kualitas belajar anak didiknya. Sehingga untuk dapat menjalankan tugasnya dengan profesional, para guru PAUD harus mengikuti diklat berjenjang untuk dapat meningkatkan kompetensinya dalam mengajar. Tiga tingkatan diklat berjenjang yang disesuaikan dengan karakteristik dan kebutuhan guru pada setiap jenjang mampu memastikan ketercapaian kompetensi profesional guru dalam memberikan pelayanan pendidikan yang handal dan bermutu.